\documentclass[%
 amsmath,amssymb,
 aps, physrev,
pra,
]{revtex4-2}

\usepackage{graphicx}% Include figure files
\usepackage{dcolumn}% Align table columns on decimal point
\usepackage{bm}% bold math
\usepackage{lineno}
\newcommand{\bg}{
\begin{eqnarray}
}
\newcommand{\ed}{
\end{eqnarray}
}

\newcommand {\foe}{
\theta_{1\,\hat{e}_1}
}
\newcommand {\fte}{
\theta_{2\,\hat{e}_1}
}
\newcommand {\foa}{
\theta_{1\,\vec{a}}
}
\newcommand {\ftb}{
\theta_{2\,\vec{b}}
}
\newcommand {\ofoa}{
\hat{\theta}_{1\,\vec{a}}
}
\newcommand {\oftb}{
\hat{\theta}_{2\,\vec{b}}
}
\newcommand {\ofta}{
\hat{\theta}_{2\,\vec{a}}
}
\newcommand {\ofob}{
\hat{\theta}_{1\,\vec{b}}
}
\begin{document}

\preprint{APS/123}

\title{\textbf{Grassmann algebraic field theory, entanglement \& fermionic hidden clocks} 
}% 

\author{Han Geurdes}
\altaffiliation{Pensioner
\\
ORCID:0000-0002-7487-1875
}%Lines break automatically or can be forced with \\
 \email{Han Geurdes: han.geurdes@proton.me}

\begin{abstract}
With Grassmann algebra as fermions in a Feynman path-integral approach to field theory, the quantum correlation can be recovered. 
This means that a quantum field of Grassmann variables can explain the entanglement. 
In turn, this agrees with Einstein's ideas of field theory and local principles. 
Moreover, Grassmann variables can refer to an empirical reality like e.g. in QCD.
Hence, a possible but still fictitious fermionic field in conjunction with the previously introduced hidden clocks as Einstein variables, likely describes this empirical reality. 
\end{abstract}

%Use showkeys class option if keyword
                              %display desired
\maketitle 
\section{\label{sec:level1}Introduction\protect}%\tableofcontents
The problem of entanglement in quantum mechanics was described long ago \cite{epr} by Einstein Podolsky and Rosen.
Einstein did not fully agree with the presentation of the paper and in 1948 he wrote a simpler paper \cite{Wirk} to illustrate the problem as he saw it.
One interesting remark in \cite{Wirk} is:
"F{\"u}r die relative 
Unabh{\"a}ngigkeit r{\"a}umlich distanter Dinge (A und B) ist die Idee characteristisch:
{\"a}ussere Beeinflussung von A hat keinen unmittelbaren Einfluss auf B; dies ist als "Prinzip der Nahewirkung" bekannt, das nur in der Feld-Theorie konsequent angewendet ist."
This implies that if a field theoretical explanation of entanglement can be found, Einstein locality (i.e. the Nahewirkung principle) is demonstrated.

In the study of spin-spin entanglement, the Bell correlation formula \cite{Bell} is an important point that connects to \cite{Wirk}. 
The formula implicitly contains the claim that it fully describes all the possibilities of Einstein's local extra hidden parameters.
Although nowadays many people believe that the issue is closed with the result found in the Aspect category of experiments, probability and statistics theory tell a different story. 
We have already presented convincing evidence that there is a methodological statistical flaw in this category of experiments \cite{Han2}, \cite{Han3},  \cite{Clock}.
Note that this point about the statistics is \emph{not} a loophole. 
It is an \emph{error} in the statistical design of the experiment and a lacuna in the underlying concept of the Bell formula. 
It can't be repaired within the same approach.

A recent paper shows the possibility of hidden clocks \cite{Clock} to explain the correlation.
Furthermore, a lacuna in CHSH was already indicated \cite{Han1} for a complete classical probabilistic CHSH inequality.

Many different authors with different explanations can be mentioned. We mention a few, \cite{Ev}, \cite{Jar},  \cite{Hess0} and  \cite{Nord}. 
The case of the "many points of view" of Everett III \cite{Ev}, is likely not an alternative to a proposed clock synchronisation mechanism.
Some characteristics of the clock proposal appear to be open to probes in empirical reality. 
It is acknowledged however that it's extremely difficult in this half-philosophy, half-physics field of study to come up with proper empirical probes.
Jarrett \cite{Jar} and Hess \cite{Hess0} both approached entanglement with time variance in Bell's formula. 
This idea received a lot of very aggressive opposition.
Concerning the Kocher-Commins experiment, viz. \cite{Koch}, referred to in \cite{Nord} the following.  If clock synchronization \cite{Clock} explains entanglement in Aspect's experiment, then the entanglement in the Kocher-Commins experiment being of the same nature as with Aspect's experiment, can be explained similarly.
To end this small sequence, we also have Bohmian mechanics \cite{Bohm} where the quantum potential can play a role in entanglement. Bohm's view mathematically originates from Madelung's hydrodynamic interpretation \cite {Mad} of quantum mechanics. 

These are but some views or experiments on entanglement. 
The present author apologises for the inevitable incomplete list of authors and approaches. The author subsequently turns to the statistics of Aspect's experiment.
%%%%%%%%%%%%%%%%%%%%%%%%
\subsection{The Aspect experiment}\label{Asp}
Let's briefly describe the key elements of Aspect's experiment. 
We have the following. 
Alice's spin (polarization) measurement instrument has a unit-length parameter vector $\vec{a}=(a_1,a_2,a_3)$. Similarly, Bob has a unit length parameter vector $\vec{b}=(b_1,b_2,b_3)$.
The two instruments are at a considerable distance from each other.

The angle $x$ between both parameter vectors is measured in the plane spanned by those parameter vectors $\vec{a}$ and $\vec{b}$. 
It is $x=\angle(\vec{a},\vec{b})$.
The direction of determining the angle is e.g. going from $\vec{a}$ to $\vec{b}$. 
We have, $\cos(x)=(\vec{a}\cdot \vec{b})$.
Alice registers a spin measurement, denoted with, $\pm 1$, given parameter vector $\vec{a}$. Bob registers a $\pm 1$ with $\vec{b}$.
When Alice and Bob both register the same spin, i.e.$(1,1)$ or $(-1,-1)$, the event is denoted by $=$. When unequal it is denoted by $\neq$. 
The event countings $N(x,=)$ and $N(x,\neq)$ are crucial in the experiment. 
Aspect hoped to find in the data, for arbitrary angle $x$, the (ideal) relation 
\bg\label{asp1}
\forall_{x\in[0,2\pi)}\,\,
\sin^2(x/2)=\frac{N(x,\neq)}{N(x,=)+N(x,\neq)}
\ed
This, Einstein data  affirmative relation, violates Kolmogorov axioms, viz. \cite{Clock} or \cite{Han2}.
I.e. the right-hand side of (\ref{asp1})  is a genuine Kolmogorov (estimate) probability. 
The left-hand is not monotone non-descending for $x\in[0,2\pi)$. It is therefore not a probability function. 
Hence, nothing follows from not finding Einstein positive data in an experiment with such statistics. 
%%%%%%%%%%%%%%%%%%%%%%%%%
\section{QCD format for entanglement correlation}\label{QCD}
\subsection {Motivation}
Given the statistical problem in Aspect's experiment \cite{Clock},  
it seems justified to wonder what the physical meaning of Bell’s correlation formula actually is.
To be more specific. When one formulates a not completely Kolmogorovian model \cite{HanOld} it is rejected because "not Kolmogorovian."
But when one \emph{demonstrates} that Aspect's experiment violates Kolmogorov axioms, then the story told is: why not have a non-Kolmogorov model? 
Or, even more weird: we don't want man-made statistics such as can be found in e.g. \cite{Han3}. 
To the present author. 
All statistics originate from man's attempts to understand nature.
Introducing such semantic difference looks like an artificial attempt to hide one's errors in the statistical setup of the experiment.

The non-Kolmogorovian probability is justified by its proponents, because of quantum mechanics. 
Despite the claim that a Hilbert space gives a "new" kind of probability e.g. \cite{Pit}, the point in section-\ref{Asp} and in \cite{Clock} and \cite{Han2}, shows that Aspect's experiment occurred in classical probability terms.  
When people are counting events to construe related probabilistic event frequencies, they are operating within the boundaries of classical probability theory.
Especially, when the law of large numbers is employed to enable an experiment without looking at a specific model in terms of density $\rho(\lambda)$ and measurement functions $A(\vec{a},\lambda)$ and $B(\vec{b},\lambda)$, viz. \cite{Bell}. 
Further, if a claim is made that Bell's formula is based on some form of quantum probability variant, then what if Einstein's locality variables are, in fact, classical (Kolmogorov) probability-based?

Another weird approach is, let's introduce symmetry in the angle $x=\angle(\vec{a},\vec{b})$. 
In the latter case, one apparently forgets that during the experiment an overseer is needed to conclude this symmetry in $x=\angle(\vec{a},\vec{b})$.  
The $x$ definition can not be changed by an overseer during the experiment without the introduction of non-locality.
When analysing the data with "symmetry" an experiment is analysed that was not performed. 
Bad statistics are in this case replaced by bad experimental setup and bad data analysis practice. 

Moreover, an inconsistency argument to save CHSH from e.g. a counter proof in \cite{Han1} is not valid anymore. 
In \cite{LoHan} there's an indication of concrete mathematical incompleteness in mathematical physics.
Why would a similar incompleteness be impossible for the CHSH argumentation?

In \cite{Wirk} Einstein argues that even though variables can not be measured simultaneously, it is not sufficient proof that a particle cannot have both at the same time. 
When one accepts that view, it changes the theoretical concept, i.e. "Damit w{\"u}rde der theoretische Rahmen der Quanten-Mechanik gesprengt."
A "thou shalt not ... " rule, therefore reflects a theoretical choice. 
It doesn't go any further than that. 

To the author, who perhaps rightfully stands alone in this, it all looks like the kind of groupthink, Freud wrote about just before WWII \cite{Freud}. 
Quantum mechanics isn't superior to other thoughts, when questioning its completeness. 
Unless, like with e.g. "consciousness" in Sartre's phenomenology \cite[page 20]{Sartre}, the way quantum mechanics operates is a vicious circle. 
To the author, it is at this point, justified to ask if there is a complementary alternative 
way to approach Einstein's Nahewirkung principle. 
%%%%%%%%%%%%%%%%%%%%%%%
\subsection{Grassmann variables}
Here we will look into the possibilities of the Grassmann algebra in field theory. 
In e.g solid state physics, Grassmann variables are employed \cite{Gu} to describe fields of electrons.
According to Berezin \cite [page 11]{Ber}: "when fermions are considered, this is an anticommutative ring or, in other terminology, a Grassmann algebra."
In the terminology found in \cite{Buchb} we may look at Grassmann numbers from $\mathbb{R}_a$ 
The point of infinite generators in \cite{Buchb} is of no great concern here. 
It must be noted that Grassmann algebra is employed in quantum chromodynamics (QCD) \cite[eq (2.2.100) and (2.2.101)]{Muta}. 
It must also be noted that QCD is associated with empirical reality.
There is, therefore, no real good reason not to use Grassmann algebra in the study of entangled spins.
In a path-integral approach to quantum chromodynamics, the
Grassmann variables represent fermions.

The anti-commutativity for  Grassmann variables $\phi$
and $\psi$ is
\bg\label{ant1}
\phi\psi + \psi\phi =0
\ed
A commutation relation is valid for a number $z\in \mathbb{C}$ and a Grassmann variable like e.g. $\psi$; $z\,\psi=\psi \,z$.
The Grassmann-Berezin integration \cite{Ber} for the $\mathbb{R}_a$ type of variables is defined by 
\bg\label {ant2}
\int d\phi \, 1 = \int d\psi\, 1=0,\,
\int d\phi\, \phi=
\int d\psi\, \psi=1
\ed
Obviously,  from the anti-commutativity it follows that $\int d\psi\, \psi^2=0$, etcetera. 
In the subsequent path-integral equation the $\int d\phi(x)\,d\psi(x) f(\phi(x),\psi(x))$ must be read as
$\int d\phi\,\left( \int d \psi\,f(\phi,\psi)\right)$.
So, for instance, $\int d\phi d\psi\,\psi = \int d\phi\, 1 = 0$.
The anti-commutativity also applies to infinitesimals, $\{d\phi \, , \psi\} = \{d\psi\,, \phi\}=0$ together with, $\{d\phi,d\psi\}=0$.
The spacetime $x\in \mathbb{R}^4$ is a kind of "index" \cite[chapter 1]{Kal} to the fields $\phi=\phi(x)$ and $\psi=\psi(x)$. When not explicitly needed the $x$ dependence will be suppressed.
%%%%%%%%%%%%%%% 
\subsection {Integral equation}
\subsubsection{Preliminaries}
Let us start by defining the action. 
We suppose a purely electromagnetic Lagrangian density, $\mathcal{L}_0=-\frac{1}{4}
F_{\mu\,\nu}F^{\mu\,\nu}$. 
This form of $\mathcal{L}_0$ is an attempt to embed the Grassmann variables in a broader physics context.  
Its validity must be researched further. 

The associated action variable then is 
\bg\label{act}
S_0 =\int d^4 x \mathcal{L}_0(x)
\ed
The $S_0$ is considered a constant in the path-integration procedure. 
Furthermore, the total action is supposed to be
$S=S_0-i\sum_{n=1}^N \psi_n\phi_n$.
The variables $\psi_n=\psi(x_n)$ and $\phi_n=\phi(x_n)$.
The $x_n$ for $n=1,2,\dots,x_N$ is a coordinate in a discretised spacetime characteristic of the path-integral. 
This also means that the terminology $-i
\sum_{n=1}^N 
\psi_n\phi_n$ follows from the delta integral,
\bg\nonumber
-i\int d^4 x
\sum_{n=1}^N \delta(x-x_n)
\psi(x)\phi(x).
\ed

Let us define next the total action $S$
\bg\label{act1}
S=S_0-i\sum_{n=1}^N \psi_n\phi_n
\ed
The partitioning giving  $\psi_n\phi_n$, agrees with the partitioning of the 
$
\prod_{j=1(x_j\neq y)}^N d\phi(x_j) d\psi(x_j)
d\phi(y) d\psi(y)$. 
Subsequently, we look at $e^{iS}$.  
This factor is employed in the generating path-integral equation. The $e^{iS_0}$ is in $\mathbb{C}$. Therefore, 
\bg\label{exp}
e^{iS}=
e^{iS_0}\prod_{n=1}^N(1+\psi_n\phi_n)
\ed
The product occurs  because
$ e^{iS}=
e^{iS_0}
\prod_{n=1}^N e^{\psi_n\phi_n}=
e^{iS_0}
\prod_{n=1}^N(1+\psi_n\phi_n)$. The Taylor series for $e^{\psi\phi}$ ends at linear terms because 
$(\psi_n\phi_n)^k=0$, and $k>1$. 

Let's furthermore define here the things we will be looking at.
Firstly, $\hat{e}_1=(1,0,0)$. The parameter vectors $\vec{a}$ and $\vec{b}$ are already given in the above sections. 
Secondly,  we have for point $y=(t,\vec{y})$, where $t$ is time and $\vec{y}$ space coordinates, the folowing abbreviation $\psi=\psi(y)$ and $\phi=\phi(y)$. Then,
\bg\label{fld}
\begin{array}{l}
\foa(y)=
a_1\psi +a_2 \phi + a_3\psi\phi
\\
\ftb(y)=
b_1\phi -b_2 \psi + b_3
\\
\foe(x_k)=\psi(x_k)=\psi_k
\\
\fte(x_k)=\phi(x_k)=\phi_k
\end{array}
\ed
For the correlation, the index vectors change from $\hat{e}_{1}$ to either $\vec{a}$ and/or $\vec{b}$.

Let us then also define $[d\phi d\psi]$ as $\prod_{n=1}^N d\phi_n\psi_n$.
It is noted that for, $k,\ell =1,2,\dots,N$, the anti-commutativity $\{d\psi_k,d\phi_\ell\}=0$ is valid. Furthermore,  we also have $\{d\psi_k,\phi_\ell\}=\{d\phi_\ell, \psi_k\}=0$, together with,
$
\{\psi_k,\phi_\ell\}=
\{\phi_k,\phi_\ell\}=
\{\psi_k,\psi_\ell\}=0
$ and, $\psi_k^2=\phi_\ell^2=0$.
This implies e.g. 
\bg\label{swaps}
d\phi_1 d\psi_1
d\phi_2 d\psi_2
=
-
d\phi_1 d\phi_2
d\psi_1 d\psi_2
=
d\phi_2 d\phi_1
d\psi_1 d\psi_2
=
-
d\phi_2 d\phi_1
d\psi_2 d\psi_1
=
d\phi_2 d\psi_2
d\phi_1 d\psi_1
\ed
It, in its turn, implies that it is always possible, for proper $n>1$ and $N$ large, to go from e.g.
\bg\nonumber 
\left(
\prod_{j=1}^n d\phi(x_j) d\psi(x_j)
\right)
d\phi(y) d\psi(y)
\left(
\prod_{j=n+1}^N d\phi(x_j)
d\psi(x_j)
\right)
\ed
to, 
\bg\nonumber
\left(
\prod_{x\neq y} d\phi(x)d\psi(x)
\right)
d\phi(y) d\psi(y).
\ed
With the above definitions, it is possible to write 
$
\prod_{n=1}^N d\phi(x_n)d\psi(x_n) =
\prod_{n=1}^N
d\fte(x_n)d\foe(x_n)
$.
%%%%%%%%%%%%%%%%%%%%%%%
\subsubsection{The generating path-integral equation}
With the use of the definitions in the previous section it is possible to write down the generating path-integral similar to QCD \cite[page 60 vv]{Abers}, \cite{Muta}. 
\bg\label{p1}
Z[J_1, J_2 ]=
\int [d\fte d\foe]
e^{iS_0}\prod_{n=1}^N
(1+{\foe}_n{\fte}_n)
\exp
\left\{
i\int
d^4x'
\foe(x')J_1(x')
+
\fte(x')J_2(x')
\right\}
\ed
In the previous formula we have ${\foe}_n= \foe(x_n)$ and ${\fte}_n= \fte(x_n)$.

To continue we start with the computation of $Z[0]$.
We have the same for the $\foe(x_n)$ and $\fte(x_n)$ integration as for the $\psi(x_n)$ and $\phi(x_n)$.
And therefore, 
\bg\label{pi1}
\int d{\fte}_n d{\foe}_n
(1+{\foe}_n{\fte}_n)=
\int d\phi d\psi
(1+\psi\phi)
=
0+1
\ed
This then implies $Z[0]=e^{iS_0}$.
And we note here that $S_0=\int d^4x"\mathcal{L}_0(x")$ does not depend on the fermionic Grassmann fields $\psi(x")$ and/or $\phi(x")$.
\subsubsection{Differential Z}
Let us now turn to the generating function \cite{Muta}. Take a single $J$ as an example. 
If we, similarly to the practice  in QCD, look at a "derivative" of $Z[J]$ we firstly have
\bg\label{}
\frac{\delta Z[J]}{\delta J(y)}
=
\lim_{\epsilon \rightarrow 0}
\left(
\frac{Z[J+\epsilon \delta(x-y)] - Z[J]}{\epsilon}
\right)
\ed
Secondly,
\bg\label{dif1}
\frac{\delta Z[J]}{\delta J(y)}
=
\lim_{\epsilon \rightarrow 0}
\frac{1}{\epsilon}
\int [d\fte d\foe] e^{iS}
\left\{
\exp\left\{
i\int d^4x'
\left(
\foe(x')(J(x') +
\epsilon \delta(x'-y))
\right)
-
i\int d^4x'\foe(x')J(x')
\right\}
\right\}
\ed
If the $\epsilon \delta(x'-y))$ integral in the exponent is performed we obtain 
\bg\label{d3}
\frac{\delta Z[J]}{\delta J(y)}
=
\int [d\fte d\foe] e^{iS}
\lim_{\epsilon \rightarrow 0}
\left(
\frac{e^{i\epsilon\foe(y)}-1}{\epsilon}
\right)
\exp
\left\{
i\int d^4x'\foe(x')J(x')
\right\}
\ed
%%%%%%%%%%%%%%%%%%%%%%
\subsubsection{Exponent integral}
One may perhaps wonder 
if the position of
$
\lim_{\epsilon \rightarrow 0}
\left(
\frac{e^{i\epsilon\foe(y)}-1}{\epsilon}
\right)$ in the integral is not "trivial" because it contains Grassmann variables. 
The integral in the exponent containing the delta is
\bg\label{expi}
i\int d^4x'
\foe(x')
\left \{
J(x')
+
i\epsilon
\delta(x'-y)
\right\}
\ed
and we may note that it is equal to 
\bg\label{ei1}
i
\left(
\int d^4x'
\foe(x')
J(x')
\right)
+
i\epsilon\foe(y)
\ed
Therefore, we can note that
$
\int d^4 x'
\foe(x')
J(x')
+
\epsilon\foe(y)
=
\epsilon\foe(y)
+
\int d^4 x'
\foe(x')
J(x')
$.
So there appears to be no problem considering the position of 
$
\lim_{\epsilon \rightarrow 0}
\left(
\frac{e^{i\epsilon\foe(y)}-1}{\epsilon}
\right)$
in the integral. 
If, for clarity, we firstly set $J=0$ and then secondly have
$
\lim_{\epsilon \rightarrow 0}
\left(
\frac{e^{i\epsilon\foe(y)}-1}{\epsilon}
\right)$ 
the outcome is uniform.
This means in fact that 
$
\left.\frac{\delta Z}{\delta J}\right|_{J=0}$ is, more precise but tedious,
$
\left.\frac{\delta_\epsilon Z}{\delta_\epsilon J}\right|_{J=0,\epsilon\rightarrow 0}$
is
\bg\nonumber 
\lim_{\epsilon \rightarrow 0}
\lim_{J\rightarrow 0}
\frac{Z[J+\epsilon\delta(x-y)]-Z[J]}{\epsilon}
\ed
It means, first the delta integral in $Z[J+\epsilon\delta(x-y)]$ is evaluated. Then, secondly, the $J=0$ is implemented. Then, thirdly, the $\epsilon$ limit is evaluated.
%%%%%%%%%%%%%%%%%
\subsubsection{Green function}
The previous implies, for a second derivative, normalized with $Z[0]$ that
\bg\label{f1}
-\frac{1}{Z[0]}
\left.
\frac{\delta^2 Z}{\delta J_2 \delta J_1}\right|_{J_1=J_2=0}
=
\int[d\fte d\foe]
\prod_{n=1}^N
\left(
1+{\foe}(x_n){\fte}(x_n)
\right)
\foa(y)\fte(y)
\ed
This expression can easily be transformed, looking at equation (\ref{swaps}), to one with $\psi$ and $\phi$.
\bg\label{f2}
-\frac{1}{Z[0]}
\left.
\frac{\delta^2 Z}{\delta J_2 \delta J_1}\right|_{J_1=J_2=0}
=
\int d\phi d\psi
(1+\psi\phi)
\left(
a_1\psi +a_2 \phi + a_3\psi\phi
\right)
\left(
b_1\phi -b_2 \psi + b_3
\right)
\ed
The right-hand integral can subsequently be
rewritten like 
\bg\label{f5}
\int d\phi d\psi
(1+\psi\phi)
\left(
\sum_{k=1}^3 a_kb_k\psi\phi
-a_1b2\psi^2+a_1b_3\psi
+
a_2b_1\phi^2+a2b_3\phi
+a_3b_1\psi\phi^2 - a_3b2\psi\phi\psi
\right)
=
\\\nonumber
\int d\phi d\psi
(1+\psi\phi)
\left(
\sum_{k=1}^3 a_kb_k\psi\phi
+
a_1b_3\psi
+
a2b_3\phi
\right)
\ed
Because $\psi^2=\phi^2=0$.
Then it is clear that:
\bg\label{f6}
\int d\phi d\psi
(1+\psi\phi)
\left(
\sum_{k=1}^3 a_kb_k
\psi\phi
+
a_1b_3\psi
+
a_2b_3\phi
\right)
=
\sum_{k=1}^3 a_kb_k
\ed
The integrals $\int d\phi d\psi \,\psi = \int d\phi d\psi\, \phi =0$ and 
$
\psi\phi
\left(
\sum_{k=1}^3 a_kb_k
\psi\phi
+
a_1b_3\psi
+
a_2b_3\phi
\right)
=0
$ 
because  
$\psi^2=\phi^2=0$ 
and
$\psi\phi +\phi\psi =0$.

And so we can conclude that 
\bg\label{f7}
-\frac{1}{Z[0]}
\left.
\frac{\delta^2 Z}{\delta J_2 \delta J_1}\right|_{J_1=J_2=0}
=
\sum_{k=1}^3 a_kb_k
\ed
And that gives the (operator) Green function  \cite[equation 2.2.52]{Muta},
\bg\label{fld}
\langle 0|
\ofoa(y)\oftb(y)
|0\rangle = \sum_{k=1}^3 a_kb_k
\ed
It must be noted that 
$\langle 0|
\ofoa(y)\ofta(y)|0\rangle=1$
because 
$
\ofoa(y)\ofta(y)\propto\psi\phi
$
in the path-integral 
and similar
$
\ofob(y)\oftb(y)
\propto \psi\phi
$.
If there are $N_A$ spacetime points with 
$
\ofoa(y)\ofta(y)
$ and $N_B$ points with 
$\ofob(y)\oftb(y)$, 
then
\bg\label{many}
\langle 0|\ofoa(y_1)\ofta(y_1)\dots\ofoa(y_{N_A})\ofta(y_{N_A})\,
\ofoa(y)\oftb(y)\,
\ofob(y_1')\oftb(y_1')\dots\ofoa(y_{N_B}')\ofta(y_{N_B}')
|0\rangle 
=
\\\nonumber 
=
1 \times 1\times \dots \,
\overbrace{1\times 1\times\dots\times 1\times}^{N_A}
\left(
\sum_{k=1}^3 a_kb_k
\right)
\overbrace{ \times 1\times 1\times\dots\times 1}^{N_B}
\,
\times \dots 1 \times 1 
\ed
With, $N_A + N_B + 1 < N$ and the considered partitioning: $\mathcal{X}=x_1, x_2, \dots, \overbrace{y_1,\dots, y_{N_A}}^{\mathcal{Y}_A},\, y,\,\, \overbrace{y_1',\dots, y_{N_B}'}^{\mathcal{Y}_B}, \dots, x_{N-1},x_N$.
Note, e.g. in $\mathcal{Y}_A$ we have $N_A=|\mathcal{Y}_A|$ and $\sum_{k=1}^3 a_k^2 =1$. Similarly for $ \mathcal{Y}_B$.
The multi points operator product in (\ref{many}) appears to represent a spin-spin entanglement experiment. It allows Einstein's separability principle of \cite{Wirk}; " ... diese Dinge eine voneinander unabhängige Existenz beanspruchen, soweit diese Dinge in verschiedenen Teilen des Raumes liegen." In $\mathcal{Y}_A$ only $\vec{a}$ information. In $\mathcal{Y}_B$ only $\vec{b}$ information. 
The time ordering, in the sense of \cite[equation 2.2.42]{Muta} is implicit.
%%%%%%%%%%%%%%%%%%%%%%%%
\section {Conclusion \& discussion}
In the paper, we tried to say something about entanglement without using Bell's formula \cite{Bell}. 
The Feynman path-integral approach to a field theory was applied to the problem of entanglement. 
The model theory was QCD.
In order to have a proper Lagrangian, the quantization of a gauge field plus the implementation of a gauge condition \cite[section 2.3.3]{Muta}, appears to be necessary. 
Further research must be performed but the $S_0$ in the action is not essential for the correlation.
In addition, with non-Abelian gauge fields it's a different story than with Abelian ones. 
Entanglement occurs however in both types of gauge fields.
 
In addition, the fictitious nature of the Grassmann field somewhat resembles the Faddeev-Popov ghosts in their predominant dependence on Grassmann variables.
More importantly, 
our result in equations (\ref{fld}) and (\ref{many}) concurs with Einstein's idea of the principle of Nahewirkung in a field theoretical explanation of entanglement. 
It also seems to agree with the hidden clocks approach.
According to the present result, these hidden clocks could very well be fermionic.
Its physical meaning must be researched further. 

Finally, we note that the path-integral machinery in QCD generates testable claims that can be verified in experiment.
Perhaps this is, somehow, also true for our computations as well. 

Concerning a question about certain rules with which to approach quantum mechanics. 
In a recent paper: \cite{Fabb}, quantum contextuality is described as  "\dots the \emph{impossibility} of assigning a value to the result of a measurement of a physical property independently of the measurement context \dots "
How would the primitive, likely fermionic, clocks of \cite{Clock} fit into that representation of quantum reality?
%%%%%%%%%%%%%%%%%%
\section*{Statements}
The author has:
\\
No conflict of interest. 
\\ 
No ethical issues.
\\ 
No funding received.
\\ 
No data associated with the paper.
%%%%%%%%%%%%%%%%%%
\bibliography{apssamp}% Produces the bibliography via BibTeX.
\end{document}